\documentclass{emulateapj}

\usepackage{graphics} 
\usepackage{amsmath} 
\usepackage{amssymb}  
\usepackage{color}
\usepackage{multirow}
\usepackage{graphicx}

\usepackage{mathtools}

\definecolor{dkgreen}{rgb}{0,0.6,0}
\definecolor{gray}{rgb}{0.5,0.5,0.5}
\definecolor{mauve}{rgb}{0.58,0,0.82}

\newcommand{\tamas}[1]{\textcolor{red}{#1 [Tamas]}}
\newcommand{\matthias}[1]{\textcolor{blue}{#1 [Matthias]}}
\newcommand{\andy}[1]{\textcolor{mauve}{#1 [Andy]}}
\newcommand{\ian}[1]{\textcolor{green}{#1 [Ian]}}

\renewcommand{\citet}[1]{\citeauthor{#1} (\citeyear{#1})}
\renewcommand{\citep}[1]{(\citeauthor{#1} \citeyear{#1})}

\begin{document}
	
		
		
		
		
		\title{Subband Image Reconstruction using Differential Chromatic Refraction}

		
        \iffalse
		\author[cs]{Matthias A. Lee}
		\author[ams,cs,pha]{Tam\'{a}s Budav\'{a}ri}
		\author[uw]{Ian S. Sullivan}
		\author[uw]{Andrew J. Connolly}
		
		\address[cs]{Dept.~of Computer Science, Johns Hopkins University, Baltimore, MD,  USA}
		\address[ams]{Dept.~of Applied Mathematics \& Statistics, Johns Hopkins University, Baltimore, MD,  USA}
		\address[pha]{Dept.~of Physics \& Astronomy, Johns Hopkins University, Baltimore, MD, USA}
		\address[uw]{Dept.~of Astronomy, University of Washington, Seattle, WA, USA}
		\else
        
        \author{
Matthias A. Lee\altaffilmark{1},
Tam\'{a}s Budav\'{a}ri\altaffilmark{2,1,3},
Ian S. Sullivan\altaffilmark{4}
Andrew J. Connolly\altaffilmark{4}
}
\altaffiltext{1}{Dept.~of Computer Science, The Johns Hopkins University, Baltimore, MD, USA}
\altaffiltext{2}{Dept.~of Applied Mathematics and Statistics, The Johns Hopkins University}
\altaffiltext{3}{Dept.~of Physics and Astronomy, The Johns Hopkins University}
\altaffiltext{4}{Dept.~of Astronomy, University of Washington, Seattle, WA, USA}

\shortauthors{Lee, Budav\'ari, Sullivan, \& Connolly}

\shorttitle{Subband Image Reconstruction using DCR}

        \fi
		
		\begin{abstract}
			Refraction by the atmosphere causes the positions of sources to depend on the airmass through which an observation was taken. This shift is dependent on the underlying spectral energy of the source and the filter or bandpass through which it is observed. Wavelength-dependent refraction within a single passband is often referred to as differential chromatic refraction (DCR). With a new generation of astronomical surveys undertaking repeated observations of the same part of the sky over a range of different airmasses and parallactic angles, DCR should be a detectable and measurable astrometric signal. In this paper we introduce a novel procedure that takes this astrometric signal and uses it to infer the underlying spectral energy distribution of a source; we solve for multiple latent images at specific wavelengths via a generalized deconvolution procedure built on robust statistics. We demonstrate the utility of such an approach for estimating a partially deconvolved image, at higher spectral resolution than the input images, for surveys such as the Large Synoptic Survey Telescope (LSST). 
		\end{abstract}

			
		

	\section{Introduction}

Modern surveys
observe the same region of the sky hundreds or thousands of times over
the course of their lifetime. Due to changes in the atmosphere,
airmass, and parallactic angle of these observations the image quality and effective throughput of
these exposures can vary with time.
Time-domain and multicolor studies need to consider these images
jointly and, therefore, require advanced methods to
combine the data to maximize their information content.
Image stacking is one of these techniques and is traditionally
accomplished by either coadding images or selecting 
a small subset of the highest quality data (often referred to as lucky imaging).  
    Coadding can include convolving images to a common resolution
    before addition \citep{lucy1992co}, combining images within a
    limited range of image qualities \citep{annis2011sdss}, stacking
    all images while weighting by signal-to-noise \citep{szalay+99} or by using a matched-filter approach \citep{zackay2017coaad}.
    Lucky imaging, in contrast, selects the exposures with the highest image quality
    but at the cost of a reduction in signal-to-noise, and hence depth of the observations.
    In each of these approaches we sacrifice information (e.g., spatial resolution or depth) for computational expediency or to optimize some aspect about the properties of the images (e.g., image quality).

    \begin{figure}
    \epsscale{1.2}
    \plotone{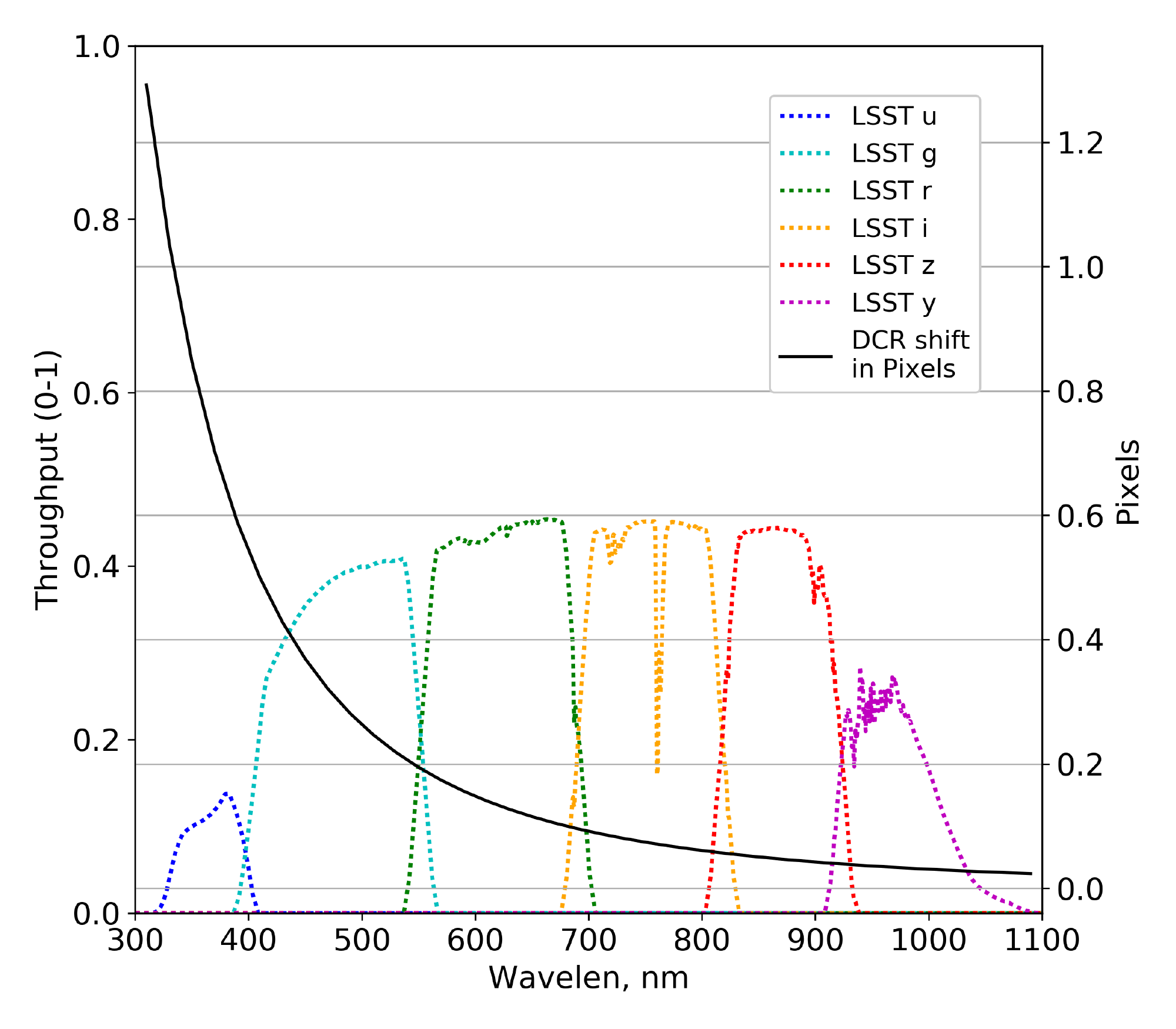}
    \caption{The DCR effect is wavelength dependent with lower
      wavelengths exhibiting a significantly larger astrometric
      offset. In this figure, we show the astrometric offset for a
      monochromatic source if we change the wavelength of the source
      by 10nm
(assuming a zenith angle of 50 degrees) on the center location of a PSF. This offset is measured in units of LSST-pixels (0.2 arcseconds). For reference we also show the LSST bandpass throughputs.}
    \label{fig:LSST_pixelshift}
    \end{figure}

	In \citet{lee2017robust}, we introduced the concept of learning an underlying model that represents the sky from a sequence of images with differing image qualities and depths. In this approach, an image $y_t$ at epoch $t$ is a simple convolution of the PSF $f_t$ and a latent model image $x$ plus some (uncorrelated) normal noise $\epsilon_{t}$,
	\begin{equation}
	y_t = f_t \ast x + \epsilon_t 
\quad\textrm{with}\quad
	\epsilon_{t,i} \sim {\mathcal{N}}\!\left(0,\sigma_{t,i}^2\right)
	\label{eq:model}
	\end{equation}
    where $\sigma_{t,i}$ is the estimated standard deviation of pixel $i$, i.e., $\sigma_t^2$ is the variance map of the $y_t$ image.
	Since the convolution is a linear operation, in the simplified limit of 1-dimensional vectors, the estimated $\hat{y}_t$ image can be written as a matrix multiplication
	\begin{equation}
	\hat{y}_t =  f_t \ast \hat{x} = F_t \hat{x}  
	\label{eq:lin}
	\end{equation}
	where $\hat{x}$ is the estimated latent image, without loss of generality.
	Given a set of $\{y_t\}$ exposures and their known $\{f_t\}$
        PSFs we have shown that we can solve for the model image $x$
        in an incremental way by iterating over the observations,
        considering them one-by-one, and updating the latent model using the data from the current epoch, $t$. Omitting the $t$ index for clarity, the multiplicative update formula is
	\begin{equation}
	\hat{x}^{\textrm{(new)}}\!= \hat{x} \odot \frac{F^T W y}{F^T W \hat{y}} 
	\label{eq:iter}
	\end{equation}
	where the dividing bar and the $\odot$ sign are elementwise division and multiplication operations, and the \mbox{$\hat{y}\!=\!F\hat{x}$} is the predicted image \cite{lee2015streaming,lee2017robust}.
	The $W$ matrix is a combination of the variance map and the
        binary masks for the exposure including censored areas, e.g.,
        saturated pixels or bad camera columns, and robust weights
        that arise as a modified minimization of the dispersion of the
        \mbox{$r_t = y_t - \hat{y}_t$} residuals in the form of a
        function $\rho(t)$ with
	\begin{equation}
	\min_{\hat{x}} \sum_{t,i} \rho\left(\frac{r_{t,i}}{\sigma_{t,i}}\right)
    \end{equation}
    rather than the traditional 
    \begin{equation}
    \min_{\hat{x}} \frac{1}{2}\sum_{t,i} \left(\frac{r_{t,i}}{\sigma_{t,i}}\right)^2 
	\end{equation}
In the classical limit of \mbox{$\rho(t)\!=\!t^2/2$}, these two
optimization problems are identical, however, for other (more) robust
$\rho$-functions, they significantly differ. The typical choice for
robust optimization is for  $\rho$ to be  quadratic for small
residuals, but in case of large deviations $\rho$ becomes a constant
value, say 1. In this way the contribution of outliers to the cost
function is limited \cite{maronna2006robust}.
%
Such robust minimization essentially corresponds to a Maximum Likelihood Estimation using a likelihood function with a longer tail than the Gaussian to accommodate outliers. 
Our standard solution is an iterative process where we re-weight the
quadratic terms with a weight derived from the robust $\rho$-function,
$\mathbb{W}(t)\!=\!\rho'(t)/t$. In the classical case of
\mbox{$\rho(t)\!=\!t^2/2$}, the weight is constant 1, but for cases
with significant outliers we use a Cauchy distribution with
$\mathbb{W}(t) = 1/(1+ t^2/c^2 )$ to down weight these outliers; see \citet{maronna2006robust,lee2017robust}.

	In this paper, we expand upon these ideas to learn not just
        the latent image but also to infer the spectral properties of
        sources from astrometric shifts introduced through
        differential chromatic refraction (DCR).
	In Section~\ref{sec:dcr} we describe our new method capable of extracting color information from sequences of monochromatic images. We apply this technique to simulated images in Section~\ref{sec:synth} where we study the limits of the approach in realistic settings. The results and potential applications of this approach are discussed in Section~\ref{sec:discsum}.

\section{Differential Chromatic Refraction}
\label{sec:dcr}
	
	Astronomical sources emit light across the continuum of the electromagnetic spectrum, $S(\lambda)$, with varying intensity. Observed broadband magnitudes or fluxes $Y$ are an integral over the filter and instrument characteristics $r(\lambda)$ through which the source is observed (including the detector's quantum efficiency, camera rotator, etc.).	

	Considering that the CCD counts photons, the equation is given by
	\begin{equation}
	Y = \frac{\int\!d\lambda\ \lambda\ r(\lambda)\,S(\lambda)}{c \int\!d\lambda\ r(\lambda) / \lambda}
	\end{equation}
	with $c$ is the speed of light.
	By substituting for a specific filter  
	\begin{equation}
	R(\lambda) = \frac{\lambda\ r(\lambda)}{c \int\!d\lambda\ r(\lambda) / \lambda} \,,
	\end{equation}
	the expression for the flux is significantly simplified,
	\begin{equation}
	Y = \int\!d\lambda\ R(\lambda)\,S(\lambda) \,.
	\end{equation}
	The above equation  applies to each pixel in a $y_t$
        exposure. The latent image is, however, complicated by the
        convolution from the PSF and the refraction of the
        atmosphere. DCR is due to the refraction of light as it passes
        through the atmosphere similar to light passing through a
        prism. The differential component of DCR refers to the change
        in the PSF across the bandpass of a given filter. It results in a positional shift and distortion of a
        source that depends on the source's spectral energy distribution
        \cite{meyers2015impact}. This shift and distortion is a
        function of wavelength, $f_t(\lambda)$, and the altitude and
        azimuth of the telescope for each epoch, $t$, of the observation.
	     \color{black}
	     In image coordinates, the effect will also depend on the orientation of the camera rotator.
	     \color{black}
        The magnitude of the
        refraction depends on the temperature and humidity of the atmosphere \cite{filippenko1982importance}. 
	
	Instead of a latent image $x$, we introduce the density image
	$\xi(\lambda)$,
	and formulate a more general model as
	\begin{equation} \label{eq:new}
	y_t  = \int\!d\lambda\ R(\lambda)\,\Big[f_t(\lambda) \ast \xi(\lambda)\Big] \ + \ \epsilon_t
	\end{equation}
	where again the $\ast$ sign denotes the spatial convolution. 
	%
	%
	In the limit of negligible dependence of the PSF on the wavelength, we get back the previous model 
	\begin{equation}
	\label{eq:approx}
	y_t = f_t \ast \left[\int\!d\lambda\,R(\lambda)\,\xi(\lambda)\right]    \ + \ \epsilon_t \ = \ f_t \ast x + \epsilon_t 
	\end{equation}
	%
	
	%
	
In Figure~\ref{fig:LSST_pixelshift} we illustrate the strength of the
astrometric shift as a function of wavelength. The black solid line
plotted over the throughput curves shows the relative
shift in pixels (0.2 arcsec) for a 10nm difference in observed wavelength. For
example the differential shift at a zenith angle of $50^\circ$ across
the extremes of the u-band, $\lambda_1=320\textrm{nm}$ and
$\lambda_2=400\textrm{nm}$, is over 6 pixels, similarly over the
g-band, $\lambda_1=400\textrm{nm}$ and $\lambda_2=560\textrm{nm}$, it
is approximately 5.5 pixels. Of course in reality, the total apparent
shift is dependent on the SED of the source
\cite{kaczmarczik2009astrometric,peters2015quasar}. The elongation of
the PSF arises because the PSF is simply the weighted summation of all
monochromatic PSFs between $\lambda_1$ and $\lambda_2$.
	


	
	To account for the wavelength dependence of DCR we, therefore,
        consider that a latent image is comprised of a number of
        sub-band images each comprising a limited wavelength interval within a given photometric passband. Given the spectral dependence of the PSF, if we can measure an astrometric offset or variation in the PSF as a function of airmass or parallactic angle (telescope pointing), we can in principle infer the underlying SED.
	Within these narrower spectral ranges, we assume a constant
        PSF and SED (though the technique can account for wavelength
        dependent PSFs even within these sub-bands).
	We  split the filter's wavelength range $[\Lambda_0, \Lambda_K)$ at preset $\{\Lambda_k : k\!=\!1\dots{}K\!-\!1\}$ values, and define the response function of these \textit{sub-bands} as
	\begin{equation}
	R_k(\lambda)= 
	\begin{cases}
	R(\lambda)& \text{if } \lambda\in[\Lambda_{k-1},\Lambda_k) \\
	\ \ 0              & \text{otherwise}
	\end{cases}
	\end{equation}
	for all $k$ from 1 to $K$. 
	Naturally,
	\begin{equation}
	y_t  = \sum_{k=1}^K \int\!d\lambda\ R_k(\lambda)\,\Big[f_t(\lambda) \ast \xi(\lambda)\Big]
	\ + \ \epsilon_t
	\end{equation}
	is equivalent to eq.(\ref{eq:new}), and its terms can be approximated similarly to that in eq.(\ref{eq:approx}), which yields
	\begin{equation}
	\hat{y}_t = \sum_k f_{t,k} \ast \hat{x}_k  = \sum_k F_{t,k}\,\hat{x}_k   
	\label{eq:sum}
	\end{equation}
	where  each $x_k$ is a latent image in the given wavelength range of sub-band $k$, and $F_{t,k}$ is the linear operator that corresponds to the convolution with the PSF $f_{t,k}$ at the appropriate wavelengths.
	

	Introducing a tall $\hat{x}$ vector that contains all these $\hat{x}_k$ (1-dimensional) images
	\begin{equation}
	\hat{x} = 
	\left[\!\!\begin{array}{c}
	\hat{x}_1 \\
	\hat{x}_2 \\
	\vdots \end{array}\!\!\right]
	\quad\textrm{and}\quad
	F_t = 
	\big[F_{t,1}\ F_{t,2}\ \dots\,\big]
	\end{equation}
	as the horizontal concatenation of the convolution matrices,
	we can verify that the equation
	\begin{equation}
	y_t = F_t \hat{x} + \epsilon_t 
	\label{eq:comp}
	\end{equation}
	is equivalent to eq.(\ref{eq:sum}). Considering that eq.(\ref{eq:comp}) is formally the same as the previous model, cf.~eqs.(\ref{eq:model}) and (\ref{eq:lin}),
	the iterative updates of eq.(\ref{eq:iter}) directly apply to our new wavelength-dependent model.
	
	Furthermore, considering that the transpose of $F$ (omitting the $t$ index) is
	\begin{equation}
	F^T = 
	\left[\!\!\begin{array}{c}
	F_1^T \\
	F_2^T \\
	\vdots \end{array}\!\!\right] \,,
	\end{equation}
	we see that the updates for the individual latent images at different $\lambda_k$ wavelengths take the familiar forms of
	\begin{align}
	\hat{x}_1^{\textrm{(new)}}\!&= \hat{x}_1 \odot \frac{F_1^T W y}{F_1^T W \hat{y}} \ , \nonumber \\ 
	\hat{x}_2^{\textrm{(new)}}\!&= \hat{x}_2 \odot \frac{F_2^T W y}{F_2^T W \hat{y}} \ , \label{eq:updates}
	\\ &\,\,\,\vdots \nonumber
	\end{align}
	and so on.
	The important difference is that now the common $\hat{y}$ term, the predicted image is the sum of the constituents wavelength dependent image,
	
	%
	\begin{equation}
	\hat{y} = \sum_k F_{k} \hat{x}_k \,.
	\end{equation}
	Note that in the limit of a single latent image, we get back the original equations and the corresponding iterative algorithm, hence this is a more general approach
	With this formalism we can use our robust iterative deconvolution method with only minor modifications.	
This iterative method is comprised of the following steps. 
We begin by initializing our models, $\big[\hat{x}_1 \dots \hat{x}_k\big]$, to be uniform in value such that the sum of all pixels is $1.0$, then start by selecting a random observation, $y_t$, for which we estimate the appropriate PSF, which we then use in conjunction with equations \ref{eq:updates}, to update our models. We repeat this procedure until we have reached convergence. Note that we reuse observations, once we have exhausted our unique supply.

\begin{figure}
\plotone{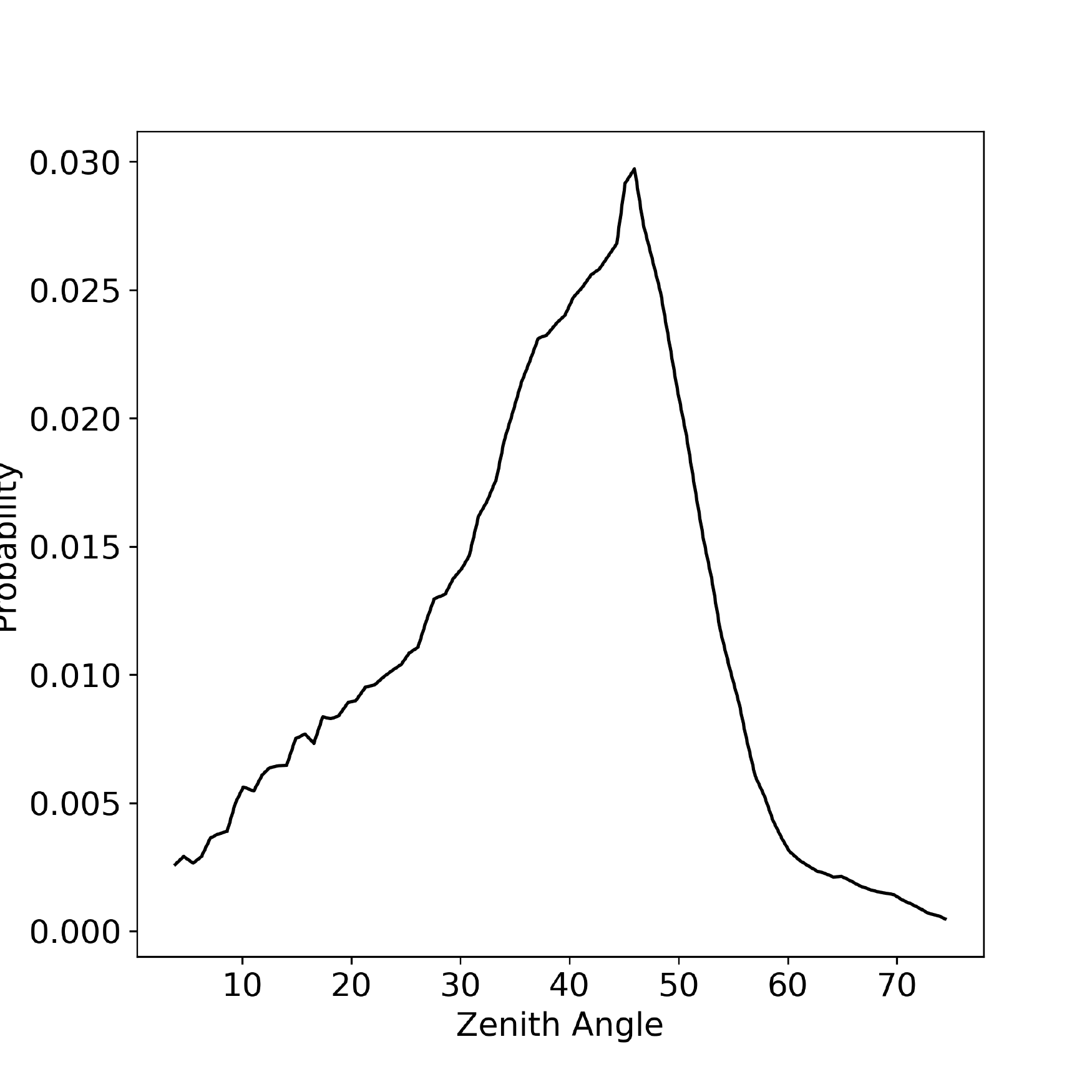}
\caption{The distribution of zenith angles sampled in our simulations, see \citet{sebag2014estimating}.}
\label{fig:zenith_angle_distribution}
\end{figure}

 \begin{figure*}
    \centering
    \includegraphics[width=0.32\textwidth]{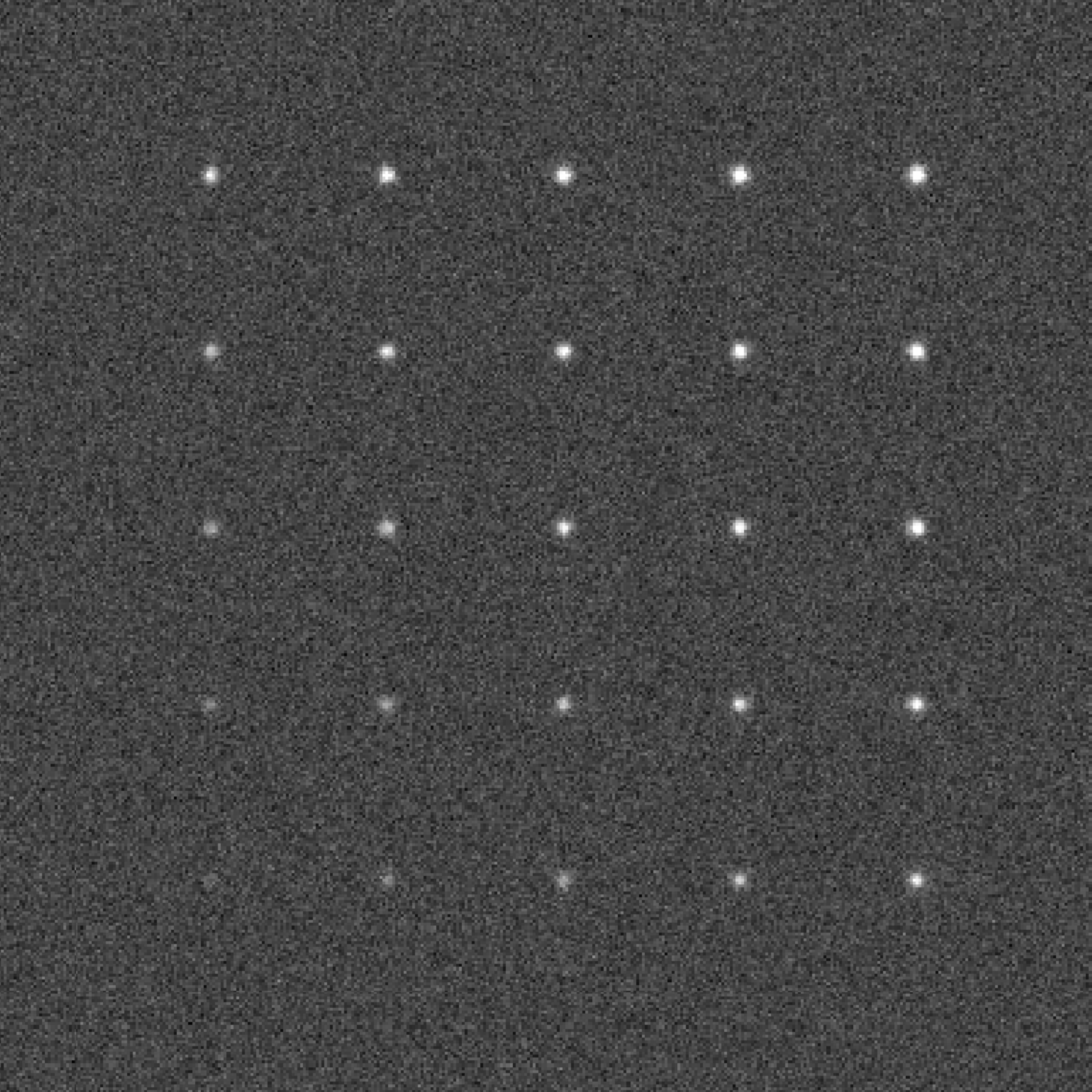}
    \includegraphics[width=0.32\textwidth]{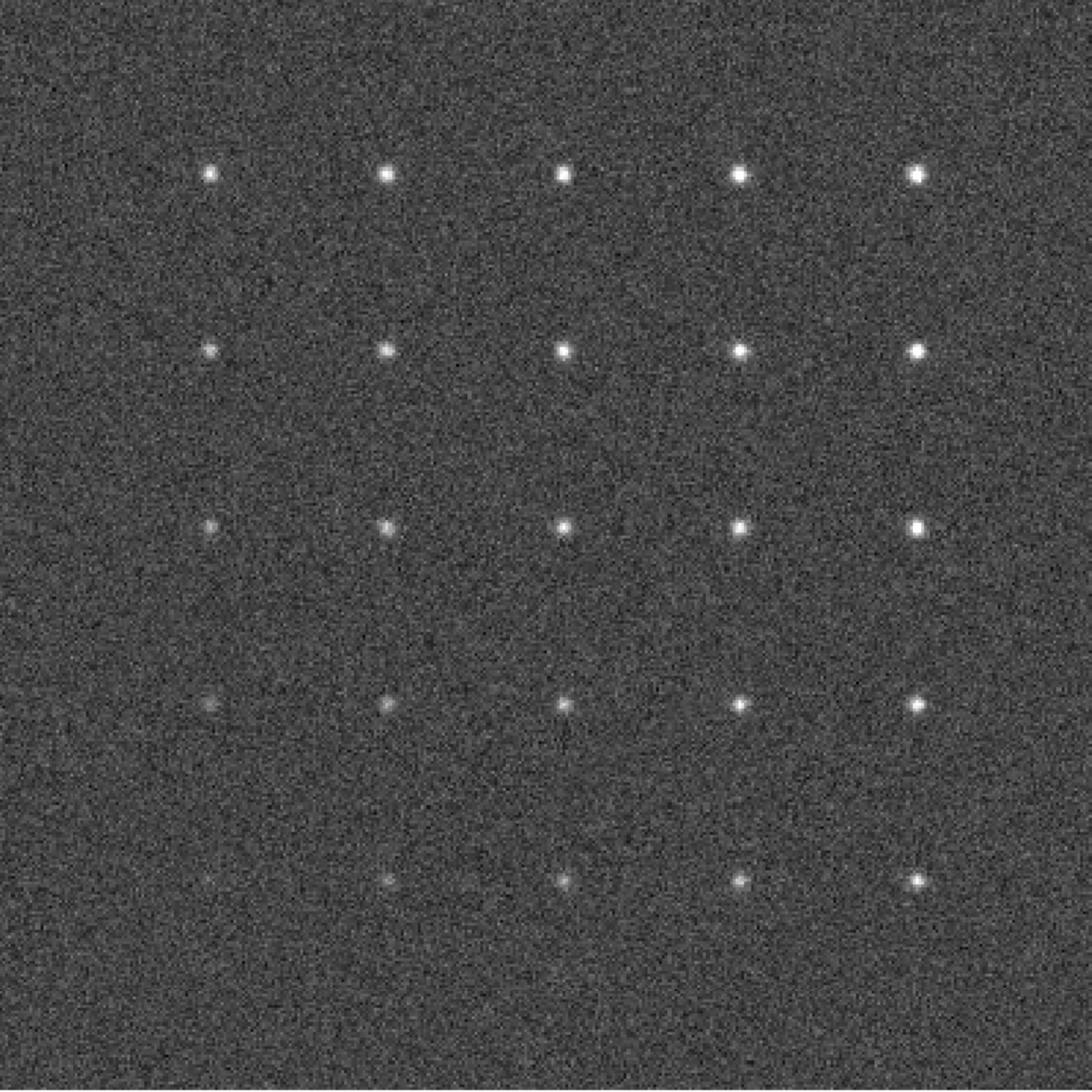}
    \includegraphics[width=0.32\textwidth]{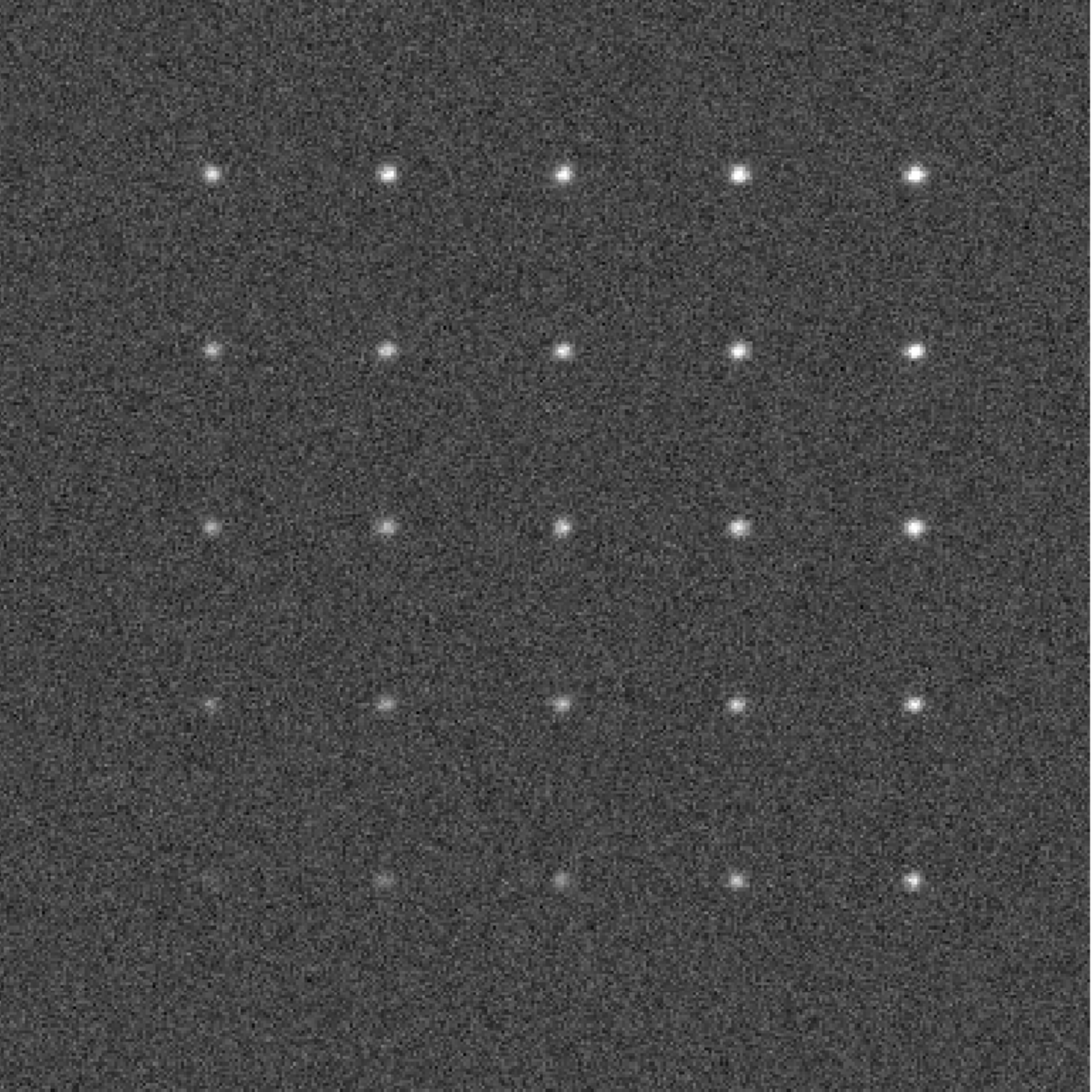}
    \caption{Synthetic exposures sampling source signal-to-noise and color. Differences between these images are slight and difficult to see by eye; \emph{left:} zenith angle
      9.85$^{\circ}$ and azimuth angle 147$^{\circ}$, source fluxes at 443.7nm and 513.5nm. \emph{center:} zenith angle 25.04$^{\circ}$ and azimuth angle 334$^{\circ}$, source fluxes at  443.7nm and 513.5nm. \emph{right:} zenith angle 50.0$^{\circ}$ and azimuth angle 77$^{\circ}$, source fluxes at 443.7nm and 513.5nm.
    }
    \label{fig:synth}
    \end{figure*}

\begin{figure*}
\centering
\includegraphics[width=0.85\textwidth,trim=2mm 5mm 5mm 8mm,clip]{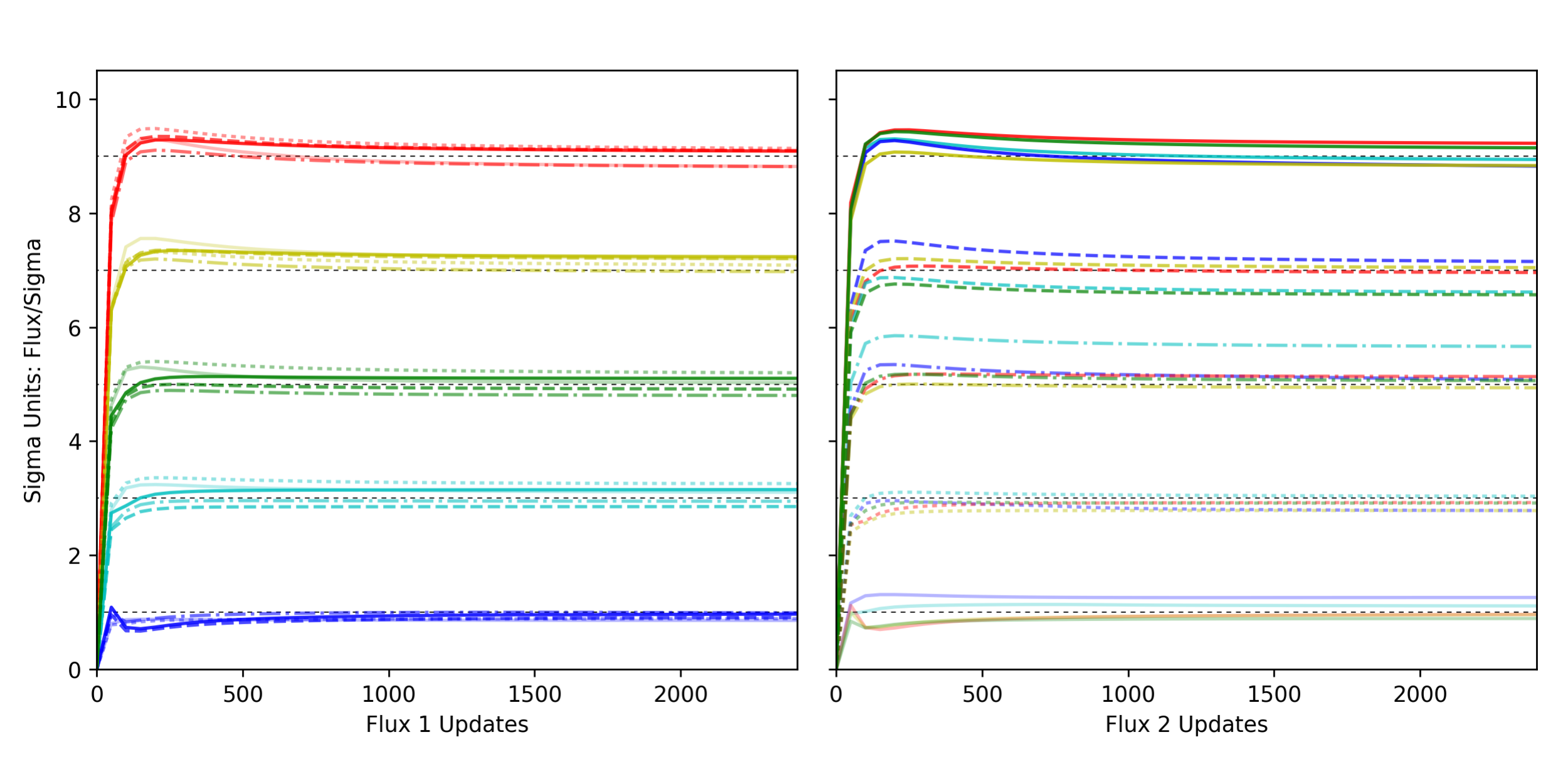}
\caption{Recovered fluxes of objects as function of iteration. \emph{Left Column:} flux 1 measured in units of $\sigma$ as a function of iteration. \emph{Right Column:} flux 2 measured in units of $\sigma$ as function of iteration. The true flux values in units of $\sigma$ for both sets of sources are  1.0, 3.0, 5.0, 7.0 and 9.0, indicated by the black horizontal lines. The flux 1 portion is represented by the color of the line and the flux 2 portion is represented by the line style. For example, the red dashed line indicates a flux 1 of $9.0\sigma$ and a flux 2 of $7.0\sigma$. Reconstruction based on a set of 50 observations. Each observation has a different realization of zenith angle and azimuth.
}
\label{fig:recovered_noisy}
\end{figure*}

	\section{Recovery from Synthetic Exposures}
	\label{sec:synth}
	
	To study the properties and limitations of this method we
        analyze synthetic images with only two components. This
        essentially corresponds to objects with spectra that only
        contain two discrete wavelengths. In other words, we attempt
        to solve for two images at preset wavelengths from a set of
        exposures that combine these wavelengths into single passband observations.
	
   \subsection{Simulated Images}
   
	Our synthetic exposures are created by observing a latent image through a PSF generated at one of two preset wavelengths. Our latent images are a simple grid of point sources where the intensity of the individual sub-bands as well as the total flux of the source vary across the image. This simulates sources with varying brightness or color. We will refer to the sub-band images as $\hat{x}_1$ and $\hat{x}_2$, corresponding to $\lambda_1$ and $\lambda_2$ respectively. 
While our simulations are of point sources, we note that this technique works equally well for resolved and even blended sources.
%
%
	
	In order to choose realistic wavelengths, we take the LSST g-band and select the effective wavelengths of the sub-bands produced by halving the g-band into two equal passbands between the wavelengths,
	\begin{equation}
	\lambda_1 = 443.7\,\textrm{nm} \quad\textrm{and}\quad\lambda_2 = 513.5\, \textrm{nm}.
	\end{equation}

	To generate a synthetic observation, we choose a zenith and azimuth angle, then generate a PSF for each wavelength at those angles and convolve each latent image with their respective PSFs.
	\color{black}
	We use the GALSIM library \citep{rowe2015galsim} to generate Kolmogorov PSFs with a FWHM of 1.0 arcseconds, the same settings used by the StarFast Simulator \citep{sullivan2016starfast}.
	\color{black}
	Finally, we add the two resulting images together into our observation. 
	It is important to correctly choose the altitude and airmass
        of the observations, as these control the strength of the DCR
        effect at  given wavelengths. The zenith angle controls the
        dispersion and the azimuth controls rotation or direction of
        the refraction. As the DCR effect is most strongly dependent
        on the zenith angle, we want to ensure we select a realistic
        set of zenith angles. To do so we randomly sample the expected
        distribution of zenith angles, generated based on LSST's OpSim \citep{lsstopsim}
        and derived from \cite{sebag2014estimating}. 
        Figure~\ref{fig:zenith_angle_distribution}  shows the probability density as a function of the zenith angle in degrees.
As expected, there is only a small fraction of the images taken at 
large angles (due to high airmass and telescope limitations).
        Typical
        zenith angles fall between $40^{\circ}$ and
        $50^{\circ}$. 
For the simulation, we sample zenith angles from this distribution.
For azimuthal angles, we randomly sample from a uniform distribution of angles ranging from $0^{\circ}$ to $360^{\circ}$.

We introduce shot noise for both the background and sources into the synthetic exposures and set the source fluxes such that their extracted measurements have realistic uncertainties. 
%
Using the StarFast Simulator of \cite{sullivan2016starfast}, we generate 200 exposures with zenith and azimuth angles, chosen as described above, using the default configuration. We run tests on subsets of these 200 observations (5, 15, and 50) to simulate how well we can recover the DCR corrected images in the early years of the survey. Additionally, when noise is introduced, we generate 50 random realizations of the noise to quantify its effect. Below we analyze these and present our results.
Figure~\ref{fig:synth} illustrates a random realization of these synthetic exposures. The sources in $\hat{x}_1$ vary from bottom to top in flux and are approximately 0, 1, 3, 5, 7 and 9 SNR, $\hat{x}_2$ has the same SNR sampling but varying from left to right.

\begin{figure*}
\centering 
\includegraphics[width=0.9\textwidth,trim=9mm 9mm 9mm 5mm,clip]{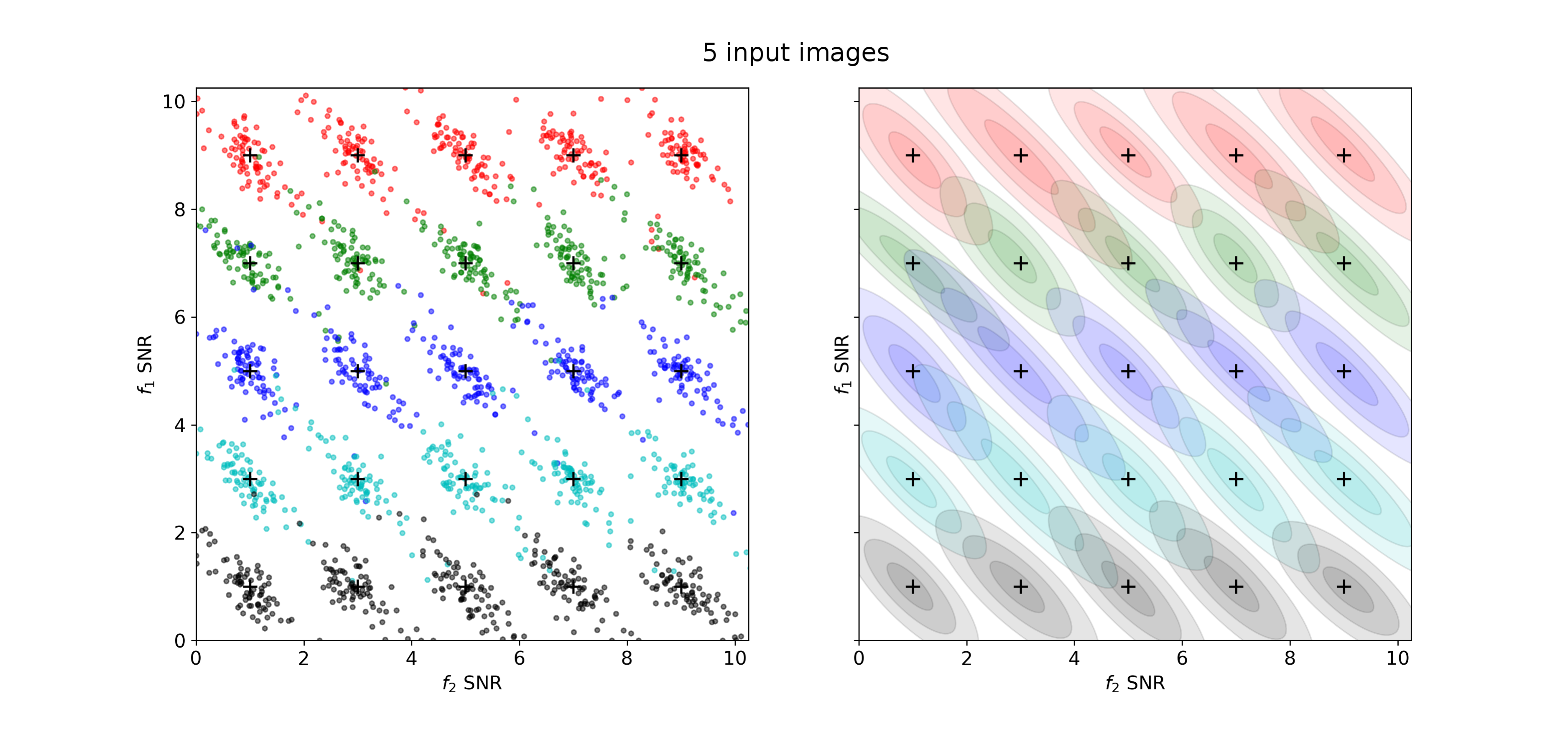}
\includegraphics[width=0.9\textwidth,trim=9mm 9mm 9mm 5mm,clip]{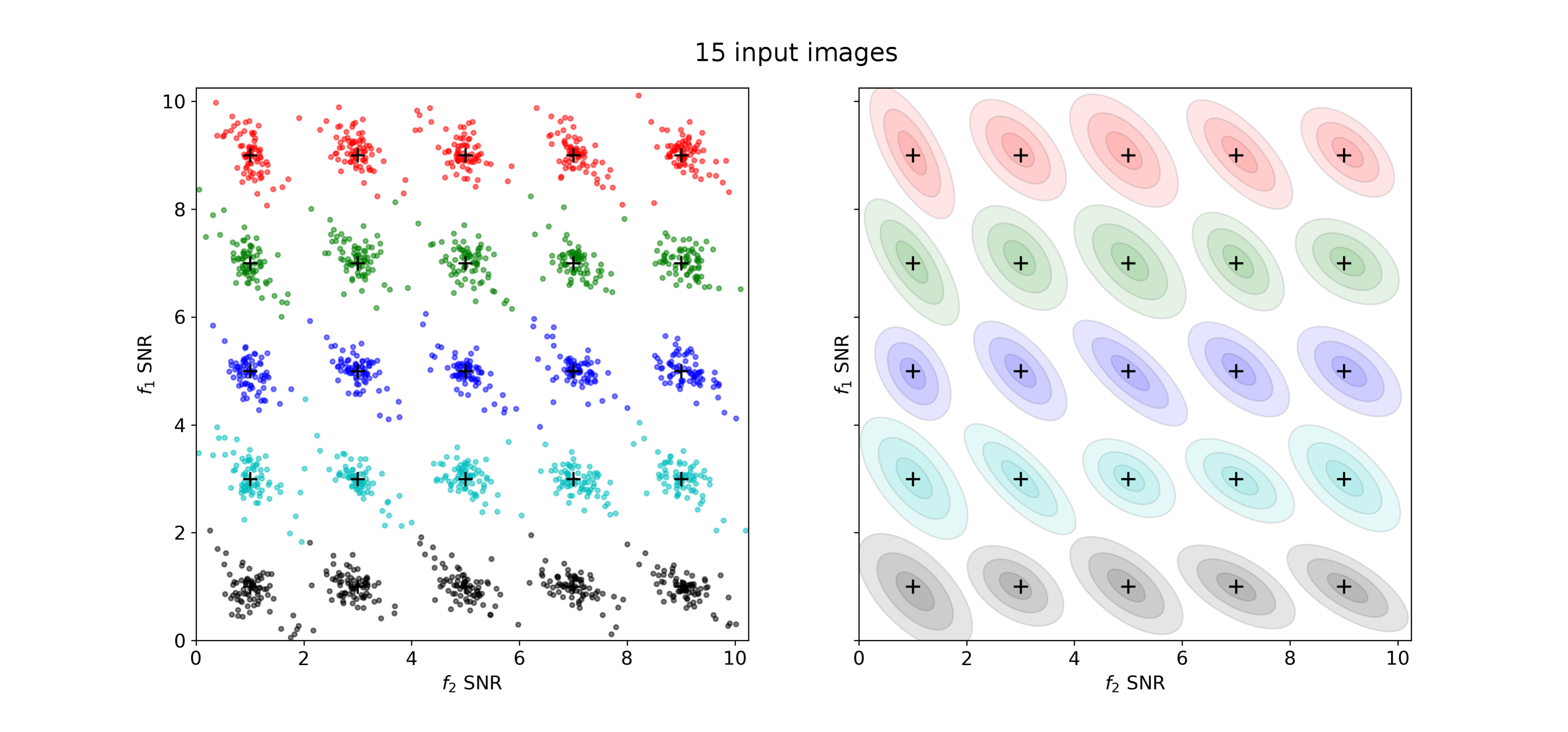}
\includegraphics[width=0.9\textwidth,trim=9mm 9mm 9mm 5mm,clip]{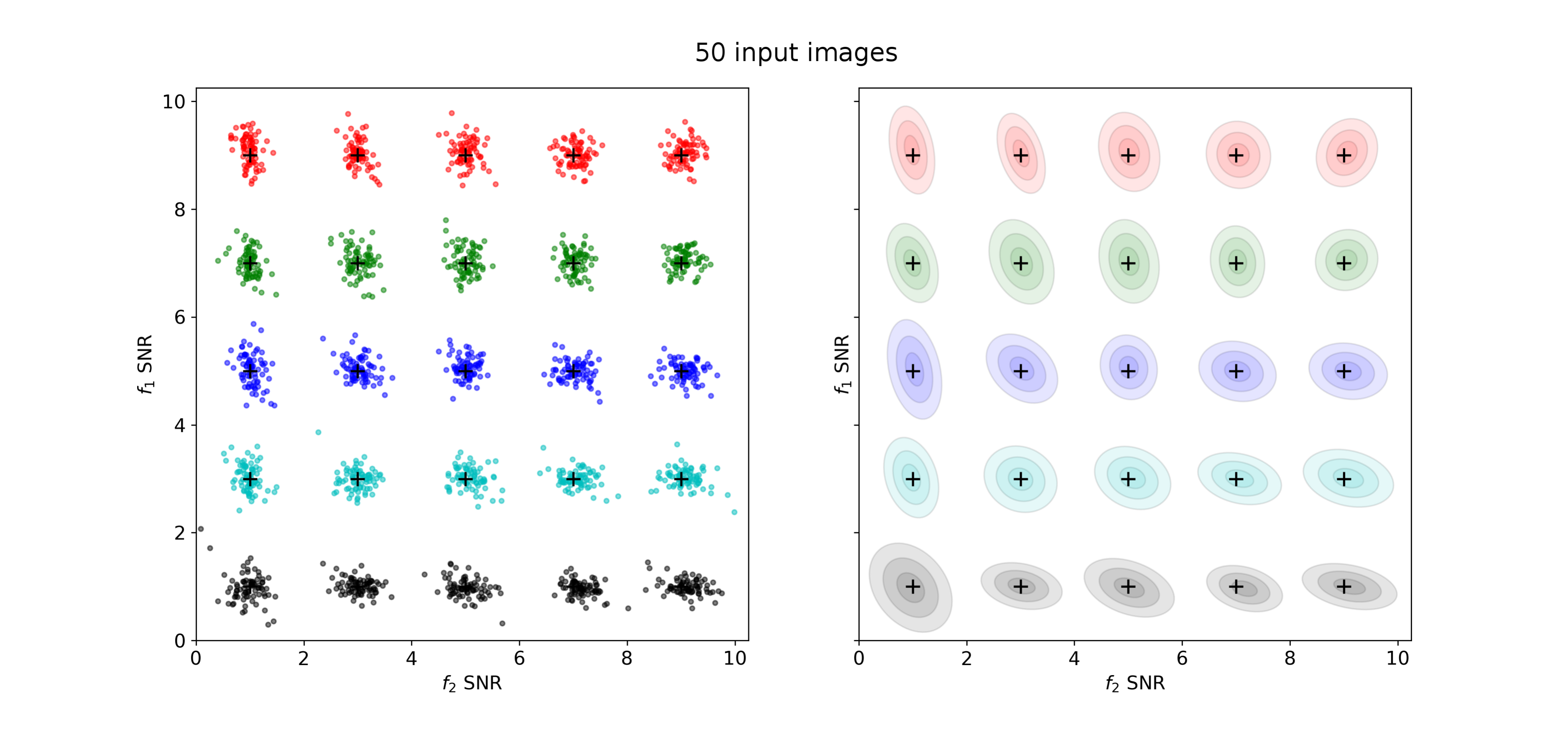}
\caption{\emph{Left:} Scatter plot of the recovered fluxes across multiple realizations; the different groups of points correspond to the different sources in Fig.~\ref{fig:synth} \emph{Right:} Covariance distribution for each flux1-flux2 pairing, ellipses drawn at $1\sigma$, $2\sigma$ and $3\sigma$. Each row corresponds to a different number of input images: \emph{first row, 5 images}; \emph{second row, 15 images}; \emph{third row, 50 images}.}
\label{fig:scatter}
\end{figure*}


\subsection{Recovered Subband Fluxes}

Our procedure is illustrated in Figure~\ref{fig:recovered_noisy} where the recovered subband fluxes are plotted for all synthetic stars as a function of the iteration assuming 50 observations.
\color{black}
For this study we measure the fluxes using simple aperture photometry.
\color{black}
The left and right panels show the fluxes in $\sigma$ units at $\lambda_1$ and $\lambda_2$, respectively.  	
By design the true fluxes are 1-, 3-, 5-, 7-, and 9-$\sigma$, which the algorithm quickly can recover despite the (intentionally) bad initialization in this illustration.
    
Due to the noise in the observed images, we see errors in the final recovered fluxes. 
To study the uncertainty of the reconstruction, we create 50 realizations and show the scatter plot of the recovered fluxes in Figure~\ref{fig:scatter}.
The left panels show the resulting realization and right panels summarize these by visualizing the sample covariance matrices.
%
The top, middle and bottom panels correspond to scenarios with 5, 15 and 50 observations, respectively.
For the small number of observations we see a strong anti-correlation between the two fluxes, as expected: their sum is better constrained than the individual subbands.
As we increase the number of observations, the errors ellipses shrink.

\begin{figure*}
\centering
\includegraphics[trim={60mm 0mm 60mm 0mm},width=0.7\textwidth]{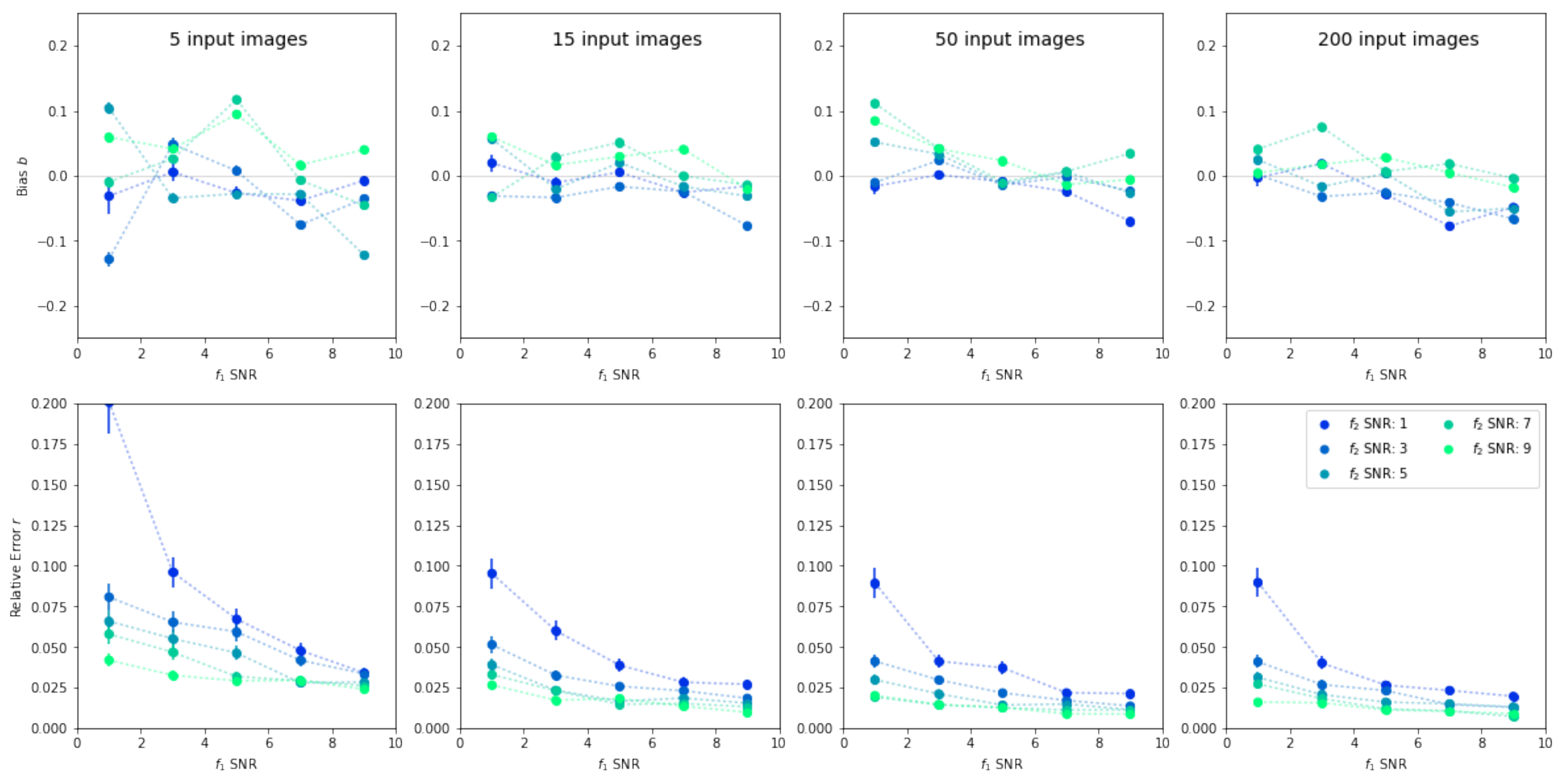}
\caption{The relative bias and uncertainty of the recovered fluxes. The flux 1 component is plotted along the x-axis and the flux 2 component is indicated by the color, \emph{see legend}. The bias and the error are reported as relative to the expected flux. The shown error bars indicate the standard error.}
\label{fig:average_flux_recov}
\end{figure*}

In Figure~\ref{fig:average_flux_recov} we study the reconstruction as function of the true flux. 
The top row shows the average bias in the flux and the bottom ones illustrate the sample variance, both of these are calculated to be relative to the expected flux. 
Beyond the initial step from 5 to 15 input images, we see the bias stay relatively constant with the increasing number of images, while the variance decreases with the signal-to-noise ratio of the sources and the number of exposures.	

%
%
%
	

\subsection{Improved Astrometry}
    
\begin{figure*}
\centering
\includegraphics[trim={60mm 0mm 60mm 0mm},width=0.7\textwidth]{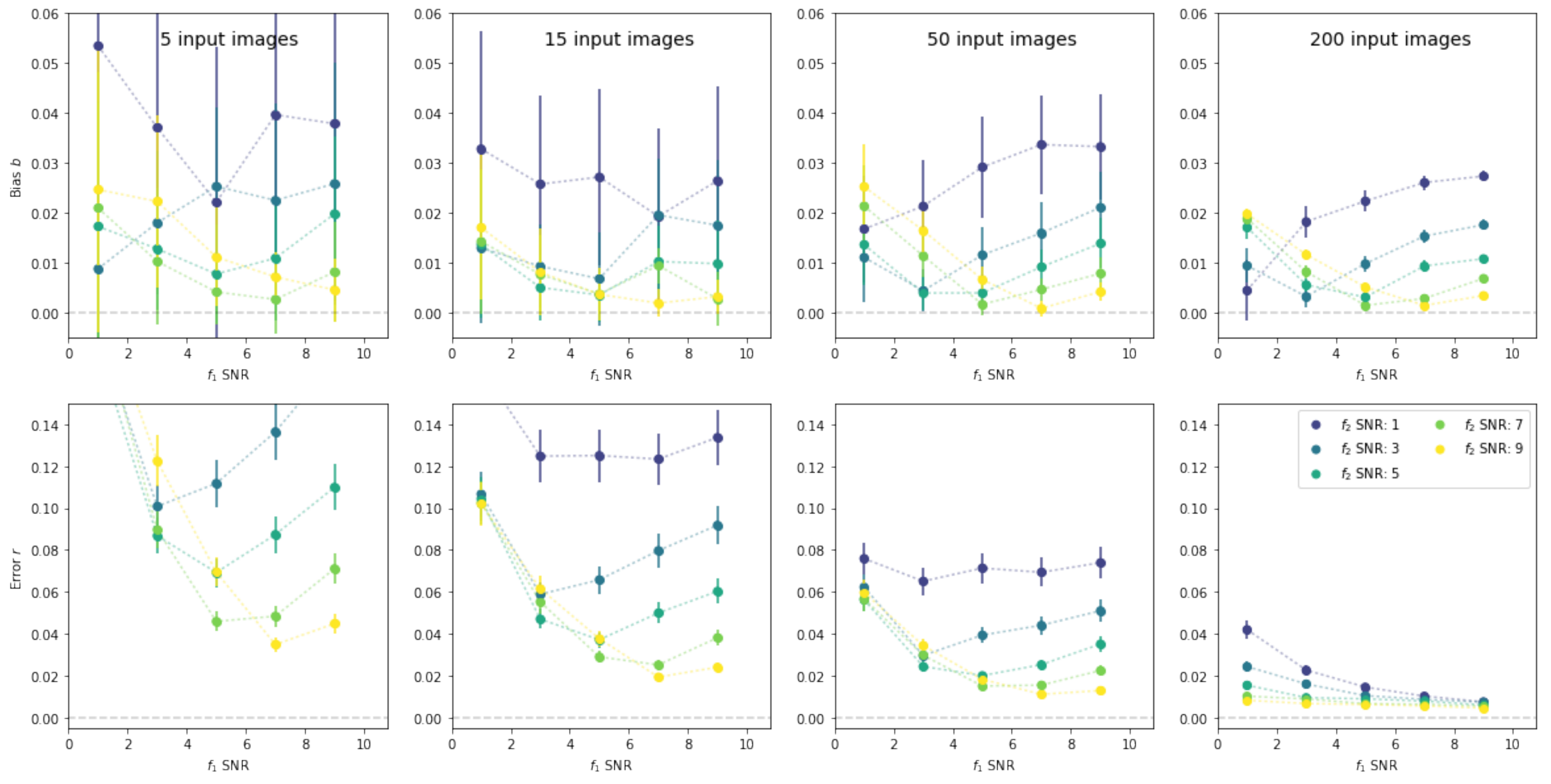}
\caption{The bias (top row) and uncertainty (bottom row) of the positions of sources measured in pixels extracted the coadded images without a correction for DCR. The flux 1 component is indicated along the x-axis of each panel and flux 2 component is represented by the color, \emph{see legend}. The shown error bars indicate the standard error.}
\label{fig:position_bias1}
\end{figure*}

Beyond evaluating the photometry, we also study the astrometric accuracy of the new reconstructions. 
For the comparison, we created coadded images by averaging the input exposures.
\color{black}
Given that the simulation did not include pointing uncertainties, the simulated exposures were not registered explicitly. Also, our study was restricted to monochromatic observations and does not make use of observed color information that multicolor surveys would have access to.
\color{black}
In Figure~\ref{fig:position_bias1} we plot the quality of the astrometry estimated for each source
\color{black}
as measured by length of the average the offset vector from the true coordinates to the observed centers of mass.
\color{black}
\color{black}
The top panels show this bias, and the bottom panels illustrate their relative errors. We see that the estimated error improves with increasing number of images, but the offset shows significant shifts even for the largest signal-to-noise scenarios. 
\color{black}
These offsets are dependent on the subband color as expected from shifts due to DCR.
	
\begin{figure*}
\centering
\includegraphics[trim={60mm 0mm 60mm 0mm},width=0.7\textwidth]{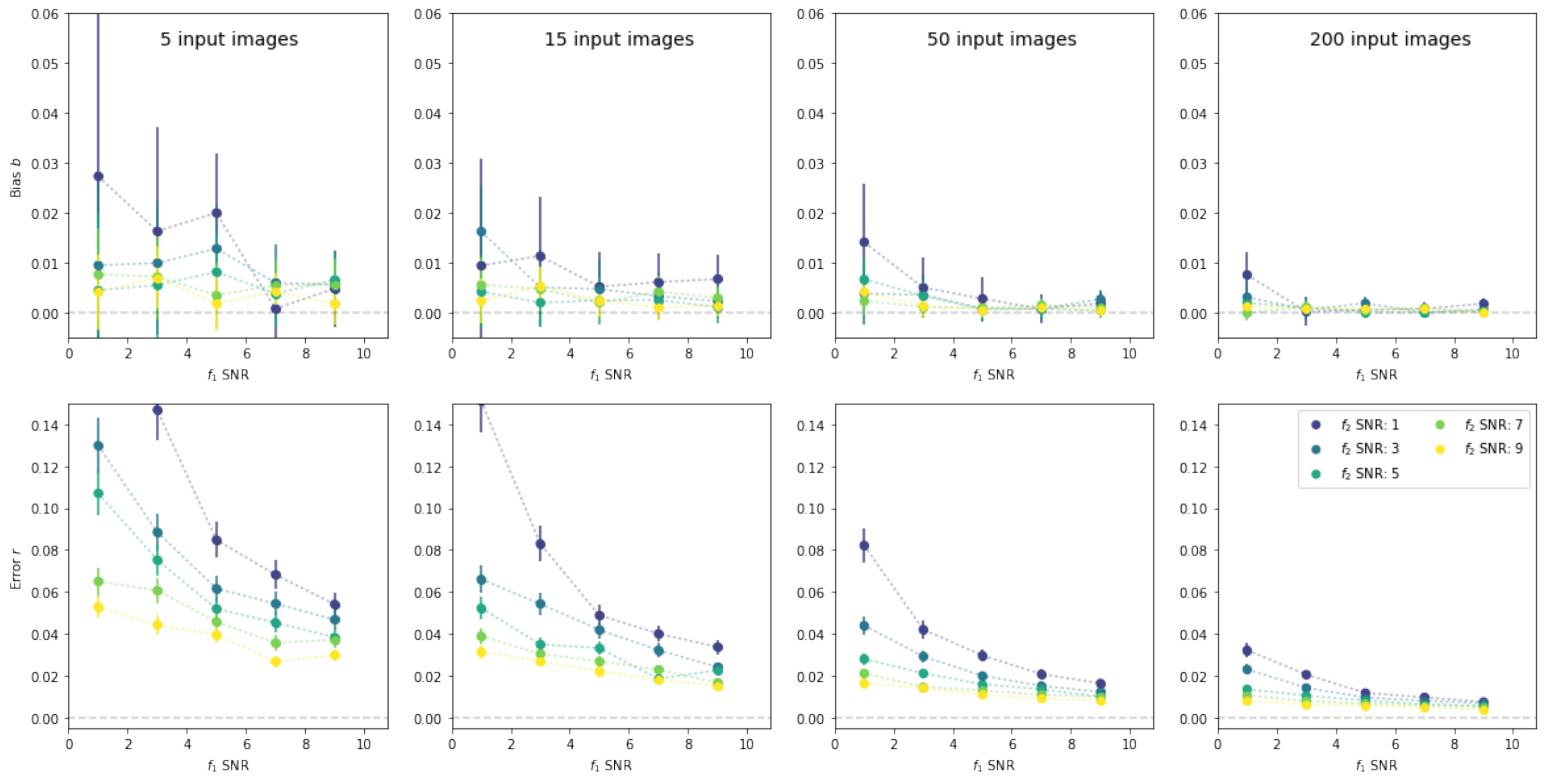}
\caption{The bias (top row) and uncertainty (bottom row) of the positions of sources measured in pixels extracted from $x_1$ and $x_2$ after the iterative image recovery. The flux 1 component is indicated along the x-axis of each panel and flux 2 component is represented by the color, \emph{see legend}. The shown error bars indicate the standard error.} 
\label{fig:position_bias2}
\end{figure*}

	
The reconstructed subband images can be combined to derive the positions of the sources and their astrometric uncertainty. 
In Figure~\ref{fig:position_bias2} the bias and error plots are shown in the same cases as the coadds. The difference is day and night: the  recovered images provide improved astrometry as we increase the signal-to-noise ratio in an subband.

In particular, the result shows little bias across the board, with an average positional bias below 0.01 pixels. Even more so, in the 50 and 200 input image case, where the average bias is below 0.005 pixels. This is largely due to the fact that the deconvolution process deblurs the source, allowing for a more accurate position finding (center of mass).
%
The error is very similar between the two methods in the case of 200 exposures, but quickly diverges for smaller numbers of input images. For 15 images, the average error across all sources for the coadd is 0.08 pixels, where as for the deconvolution result it is only half that.

\section{Discussion and Summary}
\label{sec:discsum}

Deconvolutions such as these are incredibly computationally complex, requiring multiple FFTs per input image for every loop over the set of available image. GPUs or similar massively parallel co-processors are, therefore, required in order to tackle these types workloads within reasonable timeframes. We therefore spent a considerable effort ensuring we optimize the computations our research code to take full advantage of our GPUs. While precise timings vary depending on the available hardware and the runtime settings, most of our our test runs for this article concluded in less than 60 minutes.
We make extensive use Nvidia's cuFFT library as well as a variety of custom CUDA kernels, allowing us to keep the majority of our computation on the GPU while minimizing slow transfers to and from the GPU. 

With this work, we show that small extension to our latent image recovery technique \cite{lee2017robust} enables us to account for the effects of DCR in broadband images and to use the effect of refraction to recover the underlying SED of a source, see also \citet{sullivan18}. Essentially we can use atmospheric refraction as a low resolution spectrograph to improve the spectral resolution of broadband imaging. Utilizing the LSST filter response curves for the $u$ and $g$ bands (where the DCR effect is strongest) we can improve astrometric and photometric accuracy over standard coaddition techniques that do not incorporate corrections for DCR. From the analysis of the astrometric and photometric performance we show that the improvements in uncertainties in the  recovered positions and fluxes are consistent with the increase in signal-to-noise gained by combining multiple images. For intermediate signal-to-noise sources (i.e., SNR$\sim$6) accounting for DCR in the latent image reduces the bias in the astrometric solutions by a factor of five.
%
%
%

\begin{figure}
\centering 
\includegraphics[width=0.49\textwidth,trim=15mm 5mm 5mm 5mm,clip]{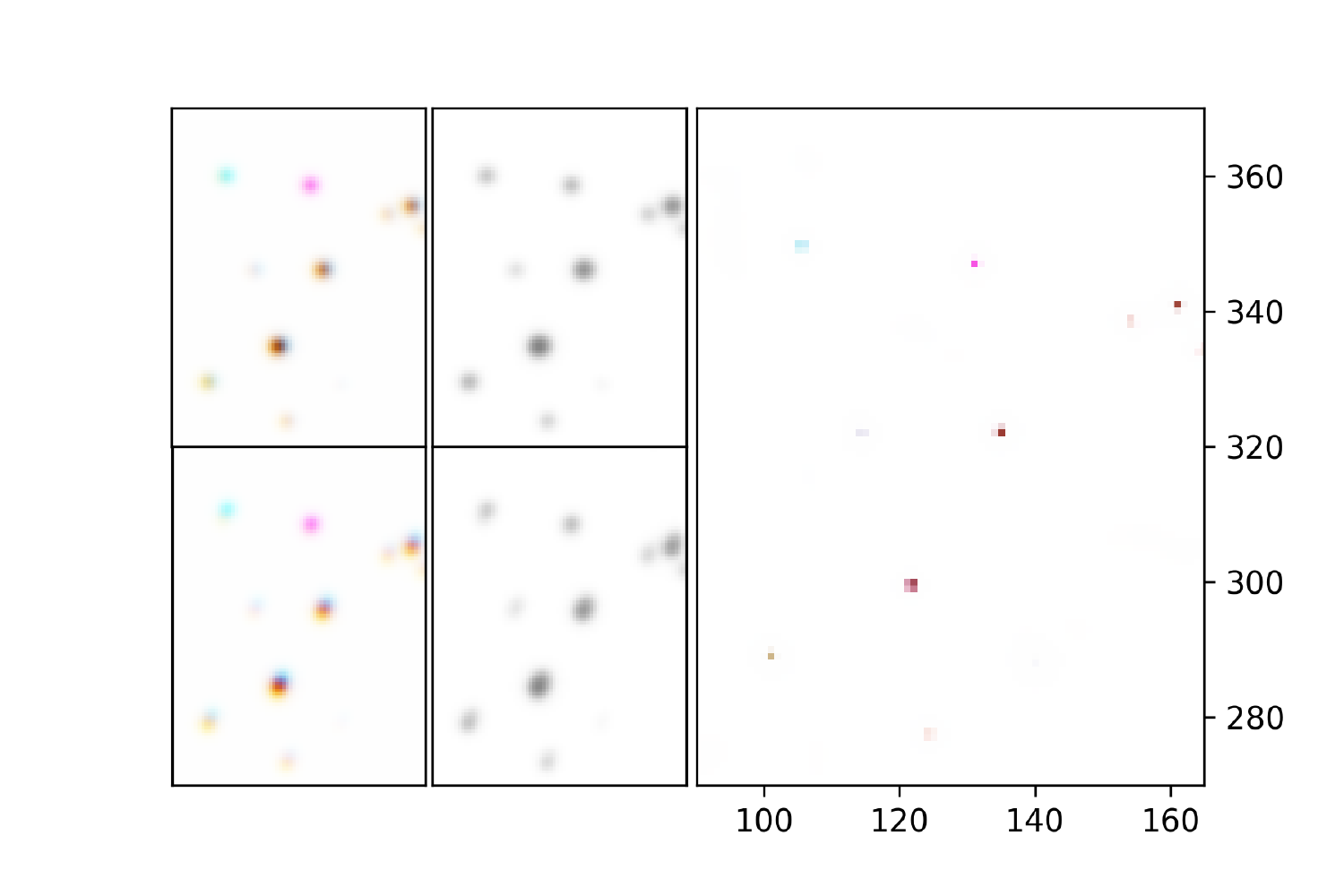}
\caption{\emph{left-column:} two typical sample observation containing a combination of 3 sub-band images plotted as an RGB image. \emph{center-column:} the corresponding simulated broadband $g$-band observations. \emph{right:} result of our deconvolution process, resolving the corresponding RGB colors of the \emph{left-column}, using only images similar to the \emph{center-column} as an input. We show a deconvolution result not only resolving a higher color resolution, than the observed images, we also significantly deblur the image and correct for the positional error caused by DCR} 
\label{fig:multi_source_sim_3band}
\end{figure}

While the main focus in this paper has been on resolving two sub-bands, this technique is easily extensible to more sub-bands. In Figure \ref{fig:multi_source_sim_3band}, we use our method to resolve 3 subbands given a realistic set of images produced using LSST's Starfast Simulator \citep{sullivan2016starfast}. In the \emph{left-column} we show two example simulated 3-band LSST images where the $g$-band was subdivided into three equally wide wavelength regions. The color dependent position of the sources and the extension of the PSF along the parallactic direction are clearly visible. The simulated observations (where we have a single broadband image) 
for the $g$-band filter and are shown in the \emph{center-column}. The DCR induced distortion to the PSF is clearly present within these images. The \emph{right} panel shows the result of processing 200 $g$-band observations, similar to those in the \emph{center-column}. This latent image resolves the sub-band images (shown as and inverted RGB) as well as partially deconvolving the underlying image. 
\color{black}
Considering that the method is able to handle PSFs and pixel sizes of various kinds simultaneously, we expect this method to work across instruments and telescopes as well.
\color{black}

\color{black}
The current model of the latent image of the sky assumes constant brightness across the observations and does not accommodate possible variability of sources explicitly. The robust statistical treatment of our approach will detect the varying flux of these objects and will essentially downweight their observation when they are significantly different from the typical brightness. Variability is a key topic that we plan to address in the future.
\color{black}

\color{black}
A caveat to current study and its conclusions is that the observations of a given astronomical object will not generally have that same distribution as the expected distribution used derived for the whole ensemble. For example, observations of the celestial pole will always occur at the same zenith angle for a given telescope. For a given object (really, a given declination), the zenith angles and parallactic angles available are also highly correlated, occupying a 1-dimensional subspace of the 2d manifold. These effect should be included for detailed analyses for specific experiment.
\color{black}

Our results are consistent with the findings of Sullivan et al.\footnote{LSST report at \url{https://dmtn-037.lsst.io}}~(2018)
and demonstrate the potential for such an approach to learn the underlying model for the sky. We note that this first in a series of papers does not include the uncertainties introduced due to PSF estimation and errors in the underlying astrometry. We will address these issues in a forthcoming paper together with the expected spectral resolution that can be achieved by surveys such as the LSST. Given the speed of the GPU implementation we expect that one year of LSST survey data for approximately 50 $\textrm{arcminute}^2$ of sky could be processed in less than 30 minutes on a single GPU.

\acknowledgements


TB acknowledges partial support from NSF Grant AST-1412566 and IIS-1447639, as well as NASA via the awards NNG16PJ23C and STScI-49721 under NAS5-26555. AJC acknowledges partial support from  DOE award DE-SC0011635, NSF awards AST-1409547 and OAC-1739419, and from the DIRAC Institute. IS acknowledges support from the NSF and the DIRAC Institute.
\color{black}
We thank the anonymous referee for the careful review and useful suggestions.
\color{black}


\newpage

	\bibliography{main.bib}

\end{document}